\documentclass[twocolumn,showpacs,preprintnumbers,amsmath,amssymb,prb,aps]{revtex4-2}
\usepackage{epsfig}
\usepackage{epstopdf}
\usepackage{graphicx}
\usepackage{dcolumn}
\usepackage{bm}
\usepackage{textcomp}
\usepackage{array}
\usepackage{caption}
\usepackage{xfrac}
\usepackage{mathtools}
\usepackage{amsmath}
\usepackage{subfigure}
\usepackage{hyperref}
\usepackage{booktabs}
\hypersetup{
        colorlinks=true,
        linkcolor=blue,
        filecolor=magenta,
        urlcolor=blue,
        pdftitle={}
        }
\begin{document}

\title{\textit{Ab-initio} study of phononic thermal conduction in ScAgC half-Heusler}
\author{Vinod Kumar Solet$^{1,}$}
\altaffiliation{vsolet5@gmail.com}
\author{Sudhir K. Pandey$^{1,}$}
\altaffiliation{sudhir@iitmandi.ac.in}
\affiliation{$^{1}$School of Mechanical and Materials Engineering, Indian Institute of Technology Mandi, Kamand - 175075, India}

\begin{abstract}
\setlength{\parindent}{4em}

In the last few years, a great deal of interest has centred on phonon studies in material science. Therefore, the present study consists of first-principles lattice calculations to comprehend the thermal expansion $\alpha(T)$ and lattice thermal conductivity $\kappa_{ph}$ of ScAgC. The obtained positive frequencies of phonon dispersion with and without non-analytic term correction (NAC) shows the dynamical stability of ScAgC in FCC structure. The estimated $\alpha(T)$ from quasi-harmonic approximation (QHA) at 300(1200) K is $\sim$4(4.6)$\times$$10^{-6}$ K$^{-1}$. The predicted value of total $\kappa_{ph}$ from phonon-phonon interaction (PPI) at 300(1200) K is $\sim$7.4(1.8) Wm$^{-1}$K$^{-1}$. The highest group velocity for acoustic $\&$ optical branches (AB $\&$ OB) is $\sim$6.7 and $\sim$3.5 km/s, respectively. The predicted average phonon lifetime ($\tau_{\lambda}$) for AB(OB) is $\sim$2.5(1.65) ps at 300 K, whereas it is $\sim$0.6(0.4) ps at 1200 K. The estimated highest heat capacity ($C_{\lambda}$) at 200 K for AB(OB) is $\sim$23.5 (19.5) meV/K. We fitted the equation $A_{\kappa}$T$^{-x_{\kappa}}$($A_{\tau}$T$^{-x_{\tau}}$) in the $\kappa_{ph}$($\tau_{\lambda}$) curve to gain a thorough understanding of temperature-dependent $\kappa_{ph}$ trend. The $x_{\tau}$ value due to total AB(OB) is estimated to be $\sim$1.04(1.02), while it is $\sim$1.03 for total branches. The $x_{\kappa}$ for total branches is calculated to be $\sim$1.02, while it is $\sim$1.04(0.95) for total AB(OB) branches, implying that AB contributes more to the total $\kappa_{ph}$. This research could be important for enhancing the properties of ScAgC regarding thermoelectric and photovoltaic applications.

\end{abstract}

\maketitle
\section{Introduction} 
\setlength{\parindent}{3em}

Energy demand is continuously growing in today’s technological world. Understanding the transport properties of materials, and in particular, the capacity to conduct heat throughout the crystal, is critical due to its technological influence on energy-related devices. The subject of heat transportation is vast in solid state physics \cite{ashcroft1976}, but we have focused attention on thermal transport by phonons here. Basically, the knowledge of phonons is very critical in accounting for many physical properties and behaviours of crystals, such as thermal transport properties, thermal expansion, phase transition, mechanical properties, and certain electrical properties (superconductivity), \textit{etc.}, \cite{ashcroft1976}. Studying the thermal behaviour of heat energy transported by the motion of atoms has always been a challenging task for researchers. The starting point for almost all theories related to lattice dynamics is harmonic approximation (HA), which ignores the concept of PPI \cite{ashcroft1976,srivastava2019}. However, HA is no longer sufficient for accessing transport properties such as $\kappa_{ph}$ due to the infinite lifetime of phonons. We can cope with this problem by taking into account the scattering of the harmonic phonons by other phonons (PPI), defects, and crystal boundaries, \textit{etc.}, \cite{ashcroft1976,chaput2013direct} because it is well known that the $\kappa_{ph}$ is a crucial property in semiconducting thermoelectrics \cite{disalvo1999thermoelectric}, solar photovoltaic (PV) cells \cite{dalal1977design}, nuclear reactors \cite{lambert1997review}, and heat management systems \cite{kim2021extremely,song2018two}, $etc$. At near or above room temperature, the PPI is an important factor to consider in the quantitative study of the $\kappa_{ph}$ of non-metallic solids, which is an anharmonic phenomenon \cite{ashcroft1976}. This anharmonicity causes complications in the prediction of these transport coefficients at a finite temperature. However, high-performance computers have contributed to some improvements in the calculation of the transport properties of solids.

In the last few decades, interesting developments toward describing the lattice dynamics and related properties of materials have been made from first-principles molecular dynamics simulations \cite{hellman2011lattice,tadano2015self}. But such calculations carry a large computational effort, and their implementation is not straightforward. In this regard, one can cope with these problems by using first-principles density functional theory (DFT) based methods, which are the appropriate and generally less expensive ways to analyse the phonon-based properties \cite{kohn1965self}. However, the use of methods beyond DFT is required for exploring the temperature-dependent transport coefficients. Recently, many-body theory based computational methods have become useful in exploring the PPI effects in crystals \cite{togo2015distributions}. Another physical property, the $\tau_{\lambda}$ of phonons, is most important to take into account in seeking the mechanism of phononic thermal conduction. The imaginary part of phonon self-energy enables us to see the strength of coupling between phonons as well as calculate the $\tau_{\lambda}$ of phonons. Nowadays, the study of anharmonic effects and the calculation of $\tau_{\lambda}$ is a highly active fields of research.

The knowledge of $\alpha(T)$ and $\kappa_{ph}$ is helpful for materials used in thermoelectric (TE) applications. TE materials are not only helpful to generate electricity from waste heat but also provide cooling power by allowing an electric current to flow through them \cite{disalvo1999thermoelectric,bell2008cooling}. These materials are particularly attractive for refrigerators, air conditioners, heat pumps, automobile applications, and \textit{etc.}, since they are reliable, quiet, and devoid of any moving parts \cite{bell2008cooling}. The most difficult challenge for researchers working with TE compounds is increasing the nondimensional parameter, \textit{figure-of-merit} $ZT$ \cite{pei2011convergence}, which is defined as $ZT$ = $S^{2}\sigma$T/$\kappa$. Where $S$, $\sigma$, $\kappa$ and T are known as the material's Seebeck coefficient, electrical conductivity, total thermal conductivity, and absolute temperature, respectively. Total $\kappa$ has two parts: an electronic part ($\kappa_{e}$) and a lattice part ($\kappa_{ph}$). The $ZT$ value of an efficient TE material should be greater than one \cite{snyder2008complex}. As a result, $ZT$ can be improved by increasing $S^{2}\sigma$ or decreasing $\kappa$. Obtaining a high $ZT$ is actually quite difficult due to the strong correlation between $S$, $\sigma$, and $\kappa_{e}$ via charge carriers \cite{ashcroft1976,sk2018exploring}. Hence, an understanding of $\kappa_{ph}$ is necessary to know the efficiency of TE materials. The $\alpha(T)$ provides the idea of a change in the length of TE materials during heating or cooling processes. In many instances, some TE materials are subjected to enough thermal or mechanical stress, which is generated during a large number of heating and cooling cycles \cite{case2012thermal}. This mechanical stress is usually referred to by the term \textquotedblleft thermal fatigue \textquotedblright. The product of elastic modulus and linear thermal expansion coefficient is an important quantity to be taken into account just before analysing the TE materials for thermal fatigue \cite{case2012thermal,music2016thermomechanical}. The lower value of this product gives the lower value of thermal fatigue in the materials. The microcracking and porosity in TE materials, which have an impact on their performance, are also caused by $\alpha(T)$ \cite{case2012thermal}. Zhang \textit{et al.} \cite{zhang2010impact} have reported the effect of microcracking on $S^{2}\sigma$ of skutterudite TE material in the 20-800 K temperature range. The $\alpha(T)$ and $\kappa_{ph}$ are directly related to the phonon calculations. Apart from electronic properties, phonon properties calculated from DFT provide reliable accuracy, which was understood in many previous works \cite{sk2022first,sk2022density,shastri2020first}. With all this in mind, the ScAgC half-Heusler (HH) compound has been chosen to explore the phonon-based properties.

Heusler compounds have recently received much interest from researchers due to their great capabilities in different energy fields, including TE   \cite{graf2011simple,shastri2020first,shastri2020thermoelectric,shiomi2011thermal}, PV solar cell \cite{yu2012identification,gruhn2010comparative}, topological insulators \cite{graf2011simple,pandey2021anab},\textit{etc.} This type of material has a general formula of $XYZ$ (HH) or $X_{2}YZ$ (full-Heuslers), where $X$ and $Y$ are mainly the transition elements and $Z$ belongs to the $p$-block element. The work of Solet \textit{et al.} predicts that ScAgC is a promising HH TE compound and the highest predicted $ZT$ is $\sim$0.53 at 1200 K temperature \cite{solet2022first}. One can achieve a high $ZT$ by lowering the $\kappa$. Further, ScAgC can also be used to make solar cell devices because it has a strong absorption of $\sim$1.7$\times$$10^{6}$ $cm^{-1}$ at photon energy $\sim$8.5 eV and a low reflectivity of $\sim$0.24 at $\sim$4.7 eV \cite{solet2022first}. At 300 K, the highest PV efficiency of $\sim$33\% is also observed at $\sim$1 $\mu m$ thickness. The remaining absorbed solar energy will be transformed into thermal energy inside the cell and could raise the temperature at the junction until the heat is not dissipated effectively to the environment \cite{royne2005cooling}. This rise in temperature reduces the mobility of charge carriers, which may be one reason for the decreased efficiency of solar cells \cite{dalal1977design}. Therefore, $\kappa_{ph}$ is critical in determining the efficiency of solar cell devices. The study of Solet \textit{et al.} \cite{solet2022first} gives full information about the electronic related properties, which is not sufficient for materials to design the TE and solar cell devices. The electronic band-gap of this compound is found to be $\sim$1.01 eV (direct band-gap) \cite{solet2022first}. Although the DFPT method was used to investigate phonon-based thermodynamical properties but $\kappa_{ph}$ was not calculated accurately in this work \cite{solet2022first}. In this direction, it is mandatory to know a more accurate $\kappa_{ph}$ in order to use ScAgC for making PV and TE devices.

Hence, the present research includes phonon-based properties estimated by first-principles calculations. First of all, phonon dispersion $\bigl($with and without including NAC$\bigl)$ and phonon DOS have been calculated by considering supercell and finite displacement approaches under HA. Both dispersion plots have no negative branches, which means that ScAgC is stable in the FCC structure. Next, $\alpha(T)$ has also been estimated via QHA. The expected room temperature value of $\alpha(T)$ is $\sim$4$\times$$10^{-6}$ K$^{-1}$, whereas it reaches a value of $\sim$4.6$\times$$10^{-6}$ K$^{-1}$ at 1200 K. Similarly, first-principle lattice calculations with an anharmonic force constant are used to capture the $\kappa_{ph}$ by assuming PPI only. The observed value of $\kappa_{ph}$ at 300 K is $\sim$7.4 Wm$^{-1}$K$^{-1}$, while it decreases to $\sim$1.8 Wm$^{-1}$K$^{-1}$ at 1200 K. Furthermore, $\tau_{\lambda}$ and $C_{\lambda}$ of each branch at different temperatures, as well as group velocity $\textbf{v}_{\lambda}$ of all phonon branches, are also calculated. The equation of temperature-dependent $\kappa_{ph}$($A_{\kappa}$T$^{-x_{\kappa}}$) and $\tau_{\lambda}$($A_{\tau}$T$^{-x_{\tau}}$) is also fitted in the respective curve. The value of $x_{\kappa}$ is found to be almost 1.02 for total branches, while it is $\sim$1.04(0.95) for AB(OB). The estimated value of $x_{\tau}$ for total branches is $\sim$1.03, whereas it is $\sim$1.04(1.02) for AB(OB), which aids in understanding the temperature-dependent $\kappa_{ph}$ trend.

\section{Computational details}

Phonon dispersion relations are carried out by means of the supercell approach and finite displacement method (FDM) \cite{kresse1995ab} in the Phonopy code \cite{togo2015first}. A 2$\times$2$\times$2 supercell is constructed to obtain displacements from the equilibrium positions of atoms in the conventional unit cell. Then forces on these supercells (with 96 atoms) are calculated from ABINIT software \cite{gonze2002first} within the projector-augmented wave (PAW) method \cite{blochl1994projector} under DFT. These atoms are displaced with a fixed harmonic distance of 0.01 \AA\ from their equilibrium positions \cite{togo2015first}. The PBE-GGA type of exchange–correlation functional has been considered in our calculations \cite{perdew1996generalized}. Converged results have been obtained by using cutoff and PAW cutoff kinetic energy for plane wave basis sets of 25 and 50 Ha, respectively. In the supercell, a 4$\times$4$\times$4 k-point grid is integrated over the Brillouin zone. The lattice parameter value of 5.6 \AA\ is used in the entire calculations \cite{solet2022first}. A force convergance criteria is set to be $5 \times 10^{-8}$ Ha/Bohr. The DFPT method, as implemented in the ABINIT code \cite{gonze2009abinit,zwanziger2012finite}, has been used to obtain born effective charges (BEC) and static dielectric constants of atoms for including long-range interactions within primitive cells. The QHA \cite{togo2015first} based method in the Phonopy code is used to evaluate the coefficient of $\alpha(T)$. Next, the phono3py \cite{togo2015distributions} package is used to calculate $\kappa_{ph}$, and a 2$\times$2$\times$2 supercell is built to obtain the second and third order force constants. But here only those supercells are considered that have interactions between only three neighbouring atoms (upto 7.5 Bohr), and the forces on these supercells have been obtained from the ABINIT code within the PAW method. Finally, these force constants are used to calculate the $\tau_{\lambda}$ from imaginary part of phonon self-energy and then $\kappa_{ph}$ by using a heavy q-mesh size of 21$\times$21$\times$21 under single mode relaxation time (SMRT) approximation \cite{togo2015distributions}.

\begin{figure}
\includegraphics[width=7.8cm, height=3.8cm]{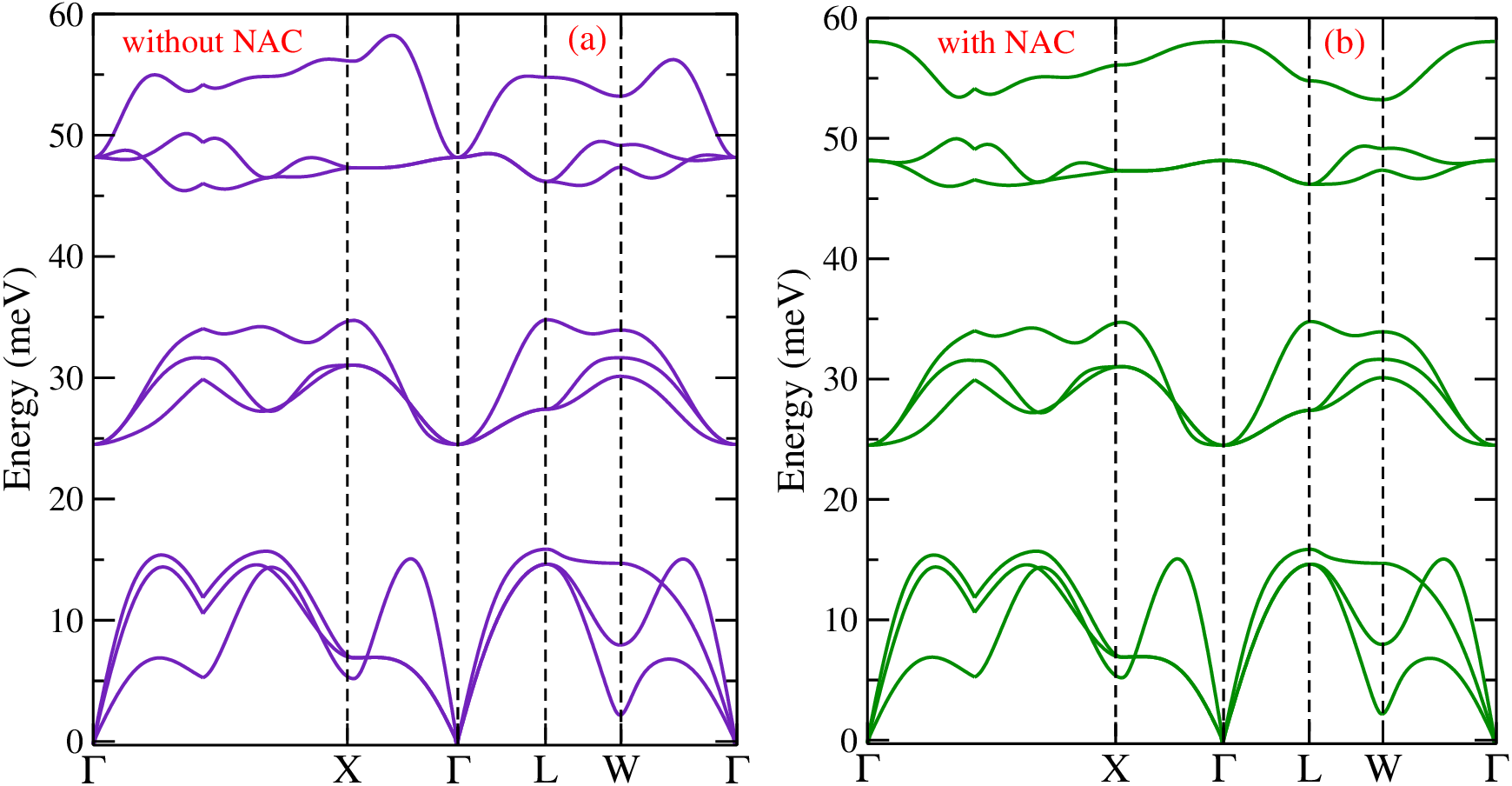} 
\caption{The harmonic phonon dispersion (a) before including NAC and (b) after including NAC of ScAgC.} 
\end{figure}

\section{Results and discussion}
\subsection{\label{sec:level2}Phonon properties}
This section presents the phonon dispersion curve and phonon DOS of ScAgC estimated under HA. To confirm the stability of this compound, the phonon dispersion is calculated in the first Brillouin zone along the high symmetry direction of $\Gamma$-$X$-$\Gamma$-$L$-$W$-$\Gamma$. The dispersion plot of Fig. 1(a) does not embody NAC. This plot contains a total of nine positive phonon branches, corresponding to three atoms in the primitive unit cell. Among them, three are AB, and the remaining six are OB. The AB have energies varying from 0 to $\sim$15.85 meV, with one being longitudinal AB (LAB) and two being transverse AB (TAB). Similarly, two longitudinal OB (LOB) and four transverse OB (TOB) are contributed in OB modes. The two AB are degenrate along the $X$-$\Gamma$ and $\Gamma$-$L$ directions. The maximum phonon energy is estimated as $\sim$58 meV. One can also observe the minimum energy gap of $\sim$8.6 meV between AB and OB. Three of the OB modes, which are located in the middle energy range ($\sim$24.5 meV to $\sim$35 meV), are well separated with an energy gap of $\sim$10 meV from the remaining three modes. These remaining branches are found in a higher energy range of $\sim$45 meV to $\sim$58 meV. Due to this gap, the coupling between the AB (OB) and OB (OB) may be weaker, which is expected to be the largest contributor to $\kappa_{ph}$. The OB modes are doubly degenerate along the $X$-$\Gamma$-$L$ direction, and they are triply degenerate at the $\Gamma$-point. The AB are almost linear near the $\Gamma$-point, implying that group velocity and phase velocity will be the same in this region \cite{ashcroft1976}. The slop of AB is used to estimate sound velocity, which is an important factor in calculating the $\kappa_{ph}$ of solids \cite{ashcroft1976}.

We proceed now to see the effect of long-range Coulomb interactions or dipole-dipole interactions on the ions present in the ionic crystals. It is well known that polar crystals become polarized by taking small atomic displacements from their equilibrium positions, and the resulting macroscopic field modifies the force constants close to the $\Gamma$-point \cite{detraux1998long}. Basically, LOB create a macroscopic electric field near the $\Gamma$-point in non-metallic solids, and thus NAC is calculated to take this contribution into harmonic phonon dispersion. As a result, the LOB is lifted up, and TOB and LOB split close to the $\Gamma$-point. One can clearly observe this LO-TO splitting at the $\Gamma$-point in Fig. 1(b). Now the maximum energy of phonons at $\Gamma$-point is $\sim$58 meV and an energy gap of $\sim$10 meV is also created. The splitting of OB also creates a minimum energy gap of $\sim$3 meV. In practice, the BEC of ions and the dielectric constant are required quantities for NAC at OB frequencies \cite{detraux1998long}. The calculated dielectric constant is $\sim$13.9, which is the same in all directions due to the cubic symmetry of ScAgC, while the BEC of Sc, Ag, and C ions is $\sim$2.5, $\sim$0.4, $\sim$ $-$2.9, respectively. According to this, the inclusion of NAC can play an important role in deciding the phonon properties. However, in the presence of NAC, the other part of the dispersion is barely affected when compared to Fig. 1(a). These theoretical aspects can be probed and verified if someone measures them by experiment.

\begin{figure}
\includegraphics[width=5.3cm, height=4.1cm]{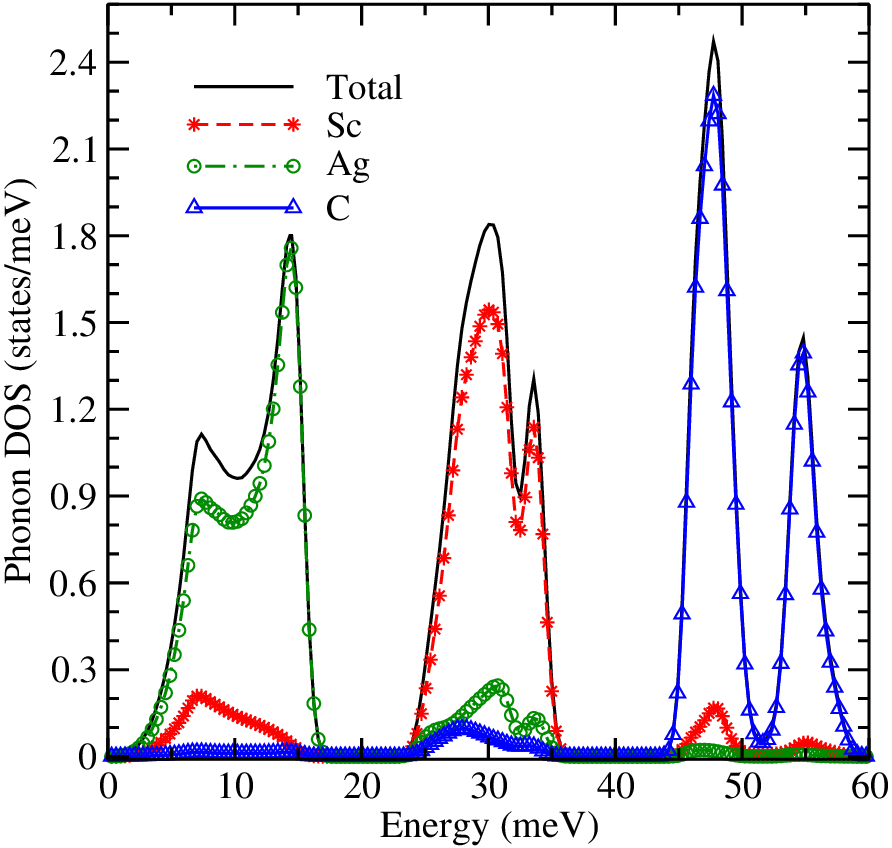} 
\caption{The total phonon DOS per unit cell and partial DOS per atom of ScAgC.} 
\end{figure}

Further, the phonon DOS and a partial DOS have been studied in order to investigate the effect of AB and OB in ScAgC during heat transfer processes. Fig. 2 indicates the obtained graph of total phonon DOS per unit cell and partial phonon DOS per atom. The total DOS plot has three main peaks around the energies of $\sim$14.5, $\sim$30 and $\sim$48 meV, respectively. One can see the gap in the DOS around 25 (40) meV, which separates the states corresponding to AB (OB) and OB (OB). In the figure, the AB in the lower energy region (below 25 meV) have the major contributions due to the vibrations of the heavier mass Ag atoms. The middle energy range of OB is mainly influenced by the atomic vibrations of Sc atoms. While the main contribution of lighter C atoms to vibrations is seen in higher (above 45 meV) energetic OB modes.    

\subsection{\label{sec:level2}Thermal expansion}

We shall now investigate the features of how the lattice can expand as a result of the thermal motion of ions at finite temperatures. The definition of $\alpha(T)$ and $\kappa_{ph}$ of compounds breaks down at the harmonic region for crystal potential. The normal-mode frequencies of purely harmonic crystals are unaffected by the change of equilibrium volume and therefore do not lead to $\alpha(T)$ \cite{ashcroft1976}. Further, anharmonic interaction in solid gives the $\alpha(T)$ because it leads to asymmetry in crystals. In this regard, it has been discovered that the QHA \cite{srivastava2019} is a respectably good approximation for capturing the $\alpha(T)$. It is known that normal mode frequencies do not always have volume dependence, but QHA considers volume-dependent phonon properties here \cite{ashcroft1976}. Accordingly, the thermal coefficient of linear expansion $\alpha(T)$ of ScAgC is estimated under QHA. In this way, the total free energy as a function of primitive cell volume is calculated at various temperatures ranging from 0 to 1200 K with a step size of 100 K, as shown in Fig. 3(a). For this, ten different supercells are created around the equilibrium lattice constant with different expansions and compressions. The total $F$ at a given temperature and volume can be estimated as, $F(T;V)$ = [$U_{el}(V)$$-$$U_{el}(V_{0})$]$+$$F_{ph}(T;V)$. Where $U_{el}(V)$$-$$U_{el}(V_{0})$ is the relative DFT energy of the electronic system, $V_{0}$ is the equilibrium volume at 0 K. $F_{ph}(T;V)$ is related to the phonon contribution to Helmholtz free energy. At each temperature, free energy has one minima corresponding to the equilibrium volume of a primitive cell, which is estimated after fitting the Birch-Murnaghan equation of states \cite{birch1947finite} to the $F$ versus volume plot. In Fig. 3(a), the solid red line connects every such energy point at equilibrium volume for a given temperature. Then, Fig. 3(b) presents these equilibrium volumes as a function of temperature up to 1200 K. The calculated equilibrium volume at 0 K is $\sim$175.7 $\AA^{3}$, while it increases to $\sim$176.5 $\AA^{3}$ at 1200 K. When compared to its ground state volume, the volume increases by up to $\sim$0.4\% at 1200 K.

\begin{figure}
\includegraphics[width=7.6cm, height=4.3cm]{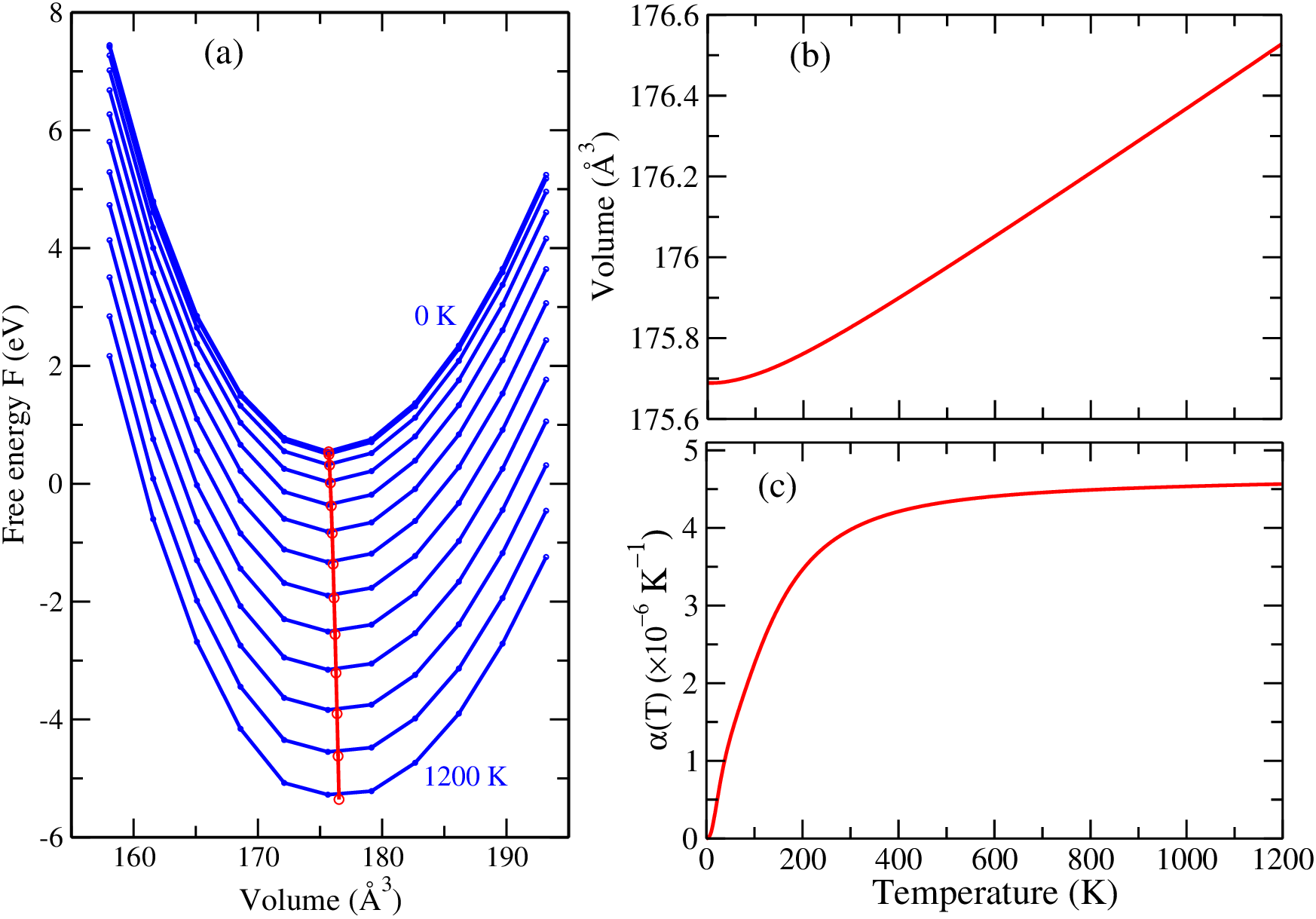} 
\caption{(a) Variation of total free energy $F$ with primitive cell volume. (b) Change in primitive cell volume with temperature. (c) The  coefficient of linear thermal expansion $\alpha(T)$ with respect to temperature for ScAgC.} 
\end{figure}

From knowing the minimum free energy for equilibrium primitive cell volume, one can easily calculate linear coefficient of thermal expansion $\alpha(T)$ from the expression of $\alpha(T)$ =  $\frac{1}{3}$$\beta(T)$ \cite{ashcroft1976}. Here, the term $\beta(T)$ is known as the volumetric thermal expansion coefficient, which is estimated as follows: $\beta(T)$ = $\frac{1}{V(T)}\frac{\partial V(T)}{\partial T}$. Where $V(T)$ represents the volume of a primitive cell as a function of temperature, as illustrated in Fig. 3(b). Because our compound has cubic symmetry, the expansion is uniform in all three directions, and thus $\alpha(T)$ is one-third of $\beta(T)$ \cite{ashcroft1976}. Fig. 3(c) shows a plot of the calculated $\alpha(T)$ versus temperature from  0-1200 K. The corresponding figure depicts a rapid increase in $\alpha(T)$ up to $\sim$200 K, followed by a slow increment in temperature up to $\sim$500 K. The rate of volume change in the crystal is highest in the 0-200 K temperature range. From $\sim$500 K to highest studied temperature, the $\alpha(T)$ shows almost constant behaviour with respect to temperature. At low temperatures, $\alpha(T)$ varies as $\sim$$T^{3}$ and becomes nearly constant at higher temperatures, exhibiting nearly the same temperature dependence behaviour to specific heat $C_{v}$ in both cases \cite{ashcroft1976,solet2022first}. The predicted value of $\alpha(T)$ at room temperature is $\sim$4$\times$$10^{-6}$ K$^{-1}$, while at 1200 K, it is $\sim$4.6$\times$$10^{-6}$ K$^{-1}$. ScAgC has lower $\alpha(T)$ values than other HH compounds like FeVSb \cite{shastri2020first}, and ZrNiSn \cite{shastri2020thermoelectric}. From an application standpoint, the information about the $\alpha(T)$ of materials is very helpful if one wants to utilize them for making real TE devices.

\subsection{\label{sec:level2}Lattice thermal conductivity}

After getting the full solution of linearized phonon Boltzmann transport equation (LBTE) using SMRT method \cite{srivastava2019}, the closed tensor form of lattice thermal conductivity $\kappa_{ph}$ is written as follows \cite{togo2015distributions}:  

\begin{eqnarray}
  \kappa_{ph} = \frac{1}{NV_{0}}\sum_{\lambda}C_{\lambda}\textbf{v}_{\lambda}\otimes \textbf{v}_{\lambda}\tau^{SMRT}_{\lambda}.
\end{eqnarray}
Here, $N$ and $V_{0}$ are the number of unit cells and volume of unit cell, respectively. $C_{\lambda}$ and $\textbf{v}_{\lambda}$ is the model specific heat and phonon group velocity of phonon mode $\lambda$, respectively. Here $\lambda$ is the phonon mode denoted by a set of (\textbf{q}, $j$) with wave vector \textbf{q} in branch $j$. $\tau^{SMRT}_{\lambda}$ is the relaxation time of corresponding phonon mode $\lambda$. $\tau^{SMRT}_{\lambda}$ is approximately considered to be phonon lifetime $\tau_{\lambda}$ in order to further calculate $\kappa_{ph}$. Then, $\tau_{\lambda}$ is obtained from the imaginary part of phonon self-energy $\Gamma_{\lambda}(\omega_{\lambda})$ by considering only PPI. The anharmonic third-order force constant is used to calculate the $\Gamma_{\lambda}(\omega_{\lambda})$, which is obtained from the many-body perturbation theory as \cite{togo2015distributions},

\begin{eqnarray}
  \Gamma_{\lambda}(\omega) = \frac{18\pi}{\hbar^{2}}\sum_{\lambda^{'}\lambda^{''}} \big| \Phi_{-\lambda\lambda^{'}\lambda^{''}}\big|^{2}\biggl\{\bigl(n_{\lambda^{'}}+ n_{\lambda^{''}}+ 1\bigl) \nonumber \\ \times \delta\bigl(\omega - \omega_{\lambda^{'}} - \omega_{\lambda^{''}}\bigl) + \bigl(n_{\lambda^{'}} - n_{\lambda^{''}}\bigl) \nonumber \\\times \Bigl[\delta\bigl(\omega + \omega_{\lambda^{'}} - \omega_{\lambda^{''}}\bigl) - \delta\bigl(\omega - \omega_{\lambda^{'}} + \omega_{\lambda^{''}}\bigl)\Bigl] \biggl\}.
\end{eqnarray}
Where $n_{\lambda}$ represents the Bose–Einstein thermal distribution function at the equilibrium of a particular phonon mode $\lambda$, which is given as, 

\begin{eqnarray}
  n_{\lambda}= \frac{1}{\exp(\hbar\omega_{\lambda}/k_{B}T)-1}
\end{eqnarray}

$\Phi_{-\lambda\lambda^{'}\lambda^{''}}$ indicates the all possible three-phonon interaction strengths between modes of $\lambda$, $\lambda^{'}$ and $\lambda^{''}$ involving in the scattering, which can be obtained from anharmonic third-order force constants.

From Eq. (2), one can calculate the $\tau_{\lambda}$ of phonon branch $\lambda$ as \cite{togo2015distributions,maradudin1962scattering},
\begin{eqnarray}
\tau^{SMRT}_{\lambda} \equiv \tau_{\lambda} = \frac{1}{2\Gamma_{\lambda}(\omega_{\lambda})}
\end{eqnarray}
Where $2\Gamma_{\lambda}(\omega_{\lambda})$ is the phonon linewidth and $\omega_{\lambda}$ denotes the harmonic phonon frequency of a mode $\lambda$.

The obtained total $\kappa_{ph}$ for ScAgC in the temperature range of 300-1200 K is presented in Fig. 4(a). One can notice the decreasing behaviour of $\kappa_{ph}$ with increasing temperature. The expected value of total $\kappa_{ph}$ at 300 K is $\sim$7.4 Wm$^{-1}$K$^{-1}$, whereas it decreases to $\sim$1.8 Wm$^{-1}$K$^{-1}$ at 1200 K. One can generally expect this decreasing behaviour since the phonon-phonon scattering rate increases with increasing temperature. The branch–dependent $\kappa_{ph}$ is also calculated to know the percentage weight of $\kappa_{ph}$ of AB and OB in the total $\kappa_{ph}$, which is shown in Fig. 4(b). In this figure, the first three AB1-AB3 are the AB, while OB1-OB6 indicate the six OB. The AB2 shows largest $\kappa_{ph}$ among all the branches and the value is $\sim$2.5(0.6) Wm$^{-1}$K$^{-1}$ at 300(1200) K. One can also observe that the AB (OB) contribute nearly 77–80 (20–23)\% of the total $\kappa_{ph}$ in the studied temperature window. This percentage decreases for AB and increases for OB as temperature rises. This is due to the fact that the number of AB (OB) modes decreases (increases) as the temperature rises. Basically, this calculation of $\kappa_{ph}$ considers only PPI. But in reality, the $\kappa_{ph}$ also affected by the phonon-electron interactions (PEI), phonon-defect interactions, \textit{etc.} Apart from this, the DFT-based phonon band structure has been used in the estimation of temperature-dependent $\kappa_{ph}$ here. However, in the real world, phonon band structure is temperature dependent. Therefore, one can get a more realistic result of $\kappa_{ph}$ with the inclusion of all the above aspects. But large computational efforts are required to address these challenges.

\begin{figure}
\includegraphics[width=8.5cm, height=4.1cm]{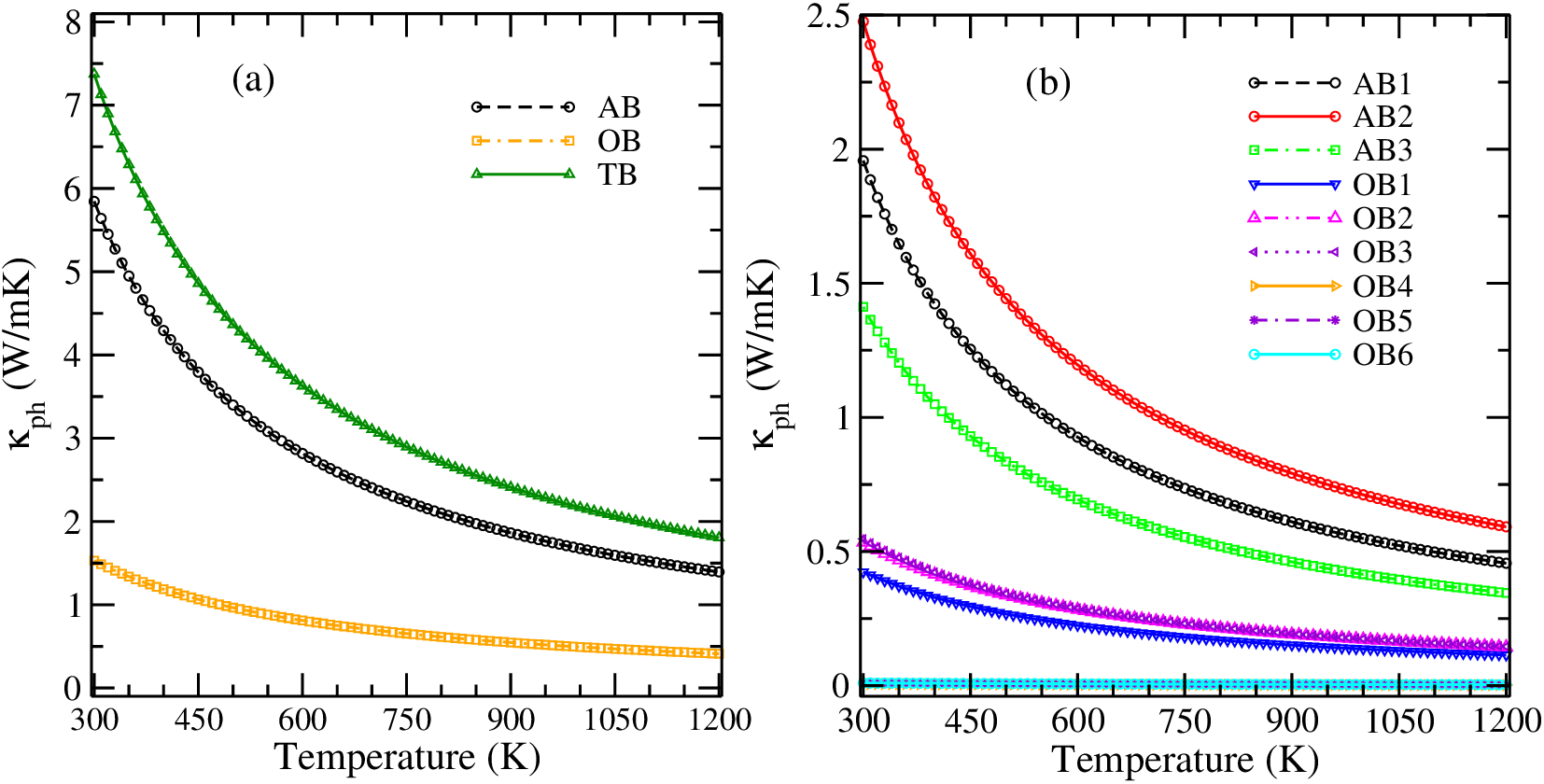} 
\caption{The calculated $\kappa_{ph}$ for (a) acoustic branches (AB), optical branches (OB), and total branches (TB) (b) nine phonon branches as a function of temperature.} 
\end{figure}

Now we shall focus on understanding the different physical parameters that contribute to the prediction of $\kappa_{ph}$. Recalling Eq. (1), the variation of $\textbf{v}_{\lambda}$ with phonon frequency, and $C_{\lambda}$, $\tau_{\lambda}$ with respect to temperature have been studied. 

The mode-dependent $C_{\lambda}$ and $\textbf{v}_{\lambda}$ can be estimated directly from the solution of eigan-value problem \cite{togo2015distributions}, 
\begin{eqnarray}
C_{\lambda} = k_{B} \Bigl(\frac{\hbar\omega_{\lambda}}{k_{B}T}\Bigl)^{2} \frac{\exp(\hbar\omega_{\lambda}/k_{B}T)}{[\exp(\hbar\omega_{\lambda}/k_{B}T)-1]^{2}},
\end{eqnarray}
and,

\begin{eqnarray}
  v_{\alpha}(\lambda)\equiv \frac{\partial\omega_{\lambda}}{\partial q_{\alpha}} \nonumber
\end{eqnarray}
\begin{eqnarray}
 = \frac{1}{2\omega_{\lambda}} \sum_{kk^{'}\beta\gamma}W_{\beta}(k,\lambda)\frac{\partial D_{\beta\gamma}(kk^{'},\textbf{q})}{\partial q_{\alpha}}W_{\gamma}(k^{'},\lambda).
\end{eqnarray}
where $\alpha$, $\beta$, and $\gamma$ are the Cartesian indices. $W_{\beta}(k,\lambda)$ is the polarization vector of $k^{th}$ atom in a unit cell, which is obtained after solving the eigan-value equation of a dynamical matrix \textbf{D(q)} \cite{togo2015distributions}.

The calculated $\textbf{v}_{\lambda}$, which is directly proportional to $\kappa_{ph}$, is presented in Fig. 5(a). In this figure, each data point represents a phonon mode for a particular q-point in the irreducible part of the Brillouin zone (IBZ). The AB3 has the highest $\textbf{v}_{\lambda}$ among all the branches, and the value is $\sim$6.7 km/s, which is almost a double value of the highest $\textbf{v}_{\lambda}$ ($\sim$3.5 km/s) of the OB9. The relatively flat dispersion of OB in Fig. 1(b) may be one of the reasons for the low $\textbf{v}_{\lambda}$ of OB compared to AB. Here, we have not considered the temperature-dependent $\textbf{v}_{\lambda}$. Nextly, the calculated $C_{\lambda}$ for all branches as a function of temperature is plotted in Fig. 5(b). In the studied temperature window, the AB have relatively large $C_{\lambda}$ compared to the OB. At 200 K, the $C_{\lambda}$ of AB1-AB3(OB1-OB3) branches is calculated to be $\sim$23.5(19.5) meV/K. At same temperature, the calculated $C_{\lambda}$ of OB4-OB5(OB6) branches is $\sim$13.5(11.5) meV/K. This can be viewed from the Bose-Einstein distribution function, in which the phonon mode population is decreased by increasing the mode's frequency at a fixed temperature. Consequently, OB have lower heat capacities, contributing less to $\kappa_{ph}$. Indeed, $C_{\lambda}$ stays almost constant with a small $\sim$2–11\% deviation ($ \sim$2\% for 630 K and $\sim$11\% for 300 K) from a constant value ($\sim$24.5 meV/K) of a classical limit of $C_{\lambda}$ at higher temperatures $\bigl($T$\gg$$\Theta_{D}$($\sim$630K \cite{solet2022first})$\bigl)$, where $\Theta_{D}$ is the Debye temperature. This observation reveals that AB modes are the dominant phonon modes and therefore make a large contribution to $\kappa_{ph}$.  

\begin{figure}
\includegraphics[width=8cm, height=4.1cm]{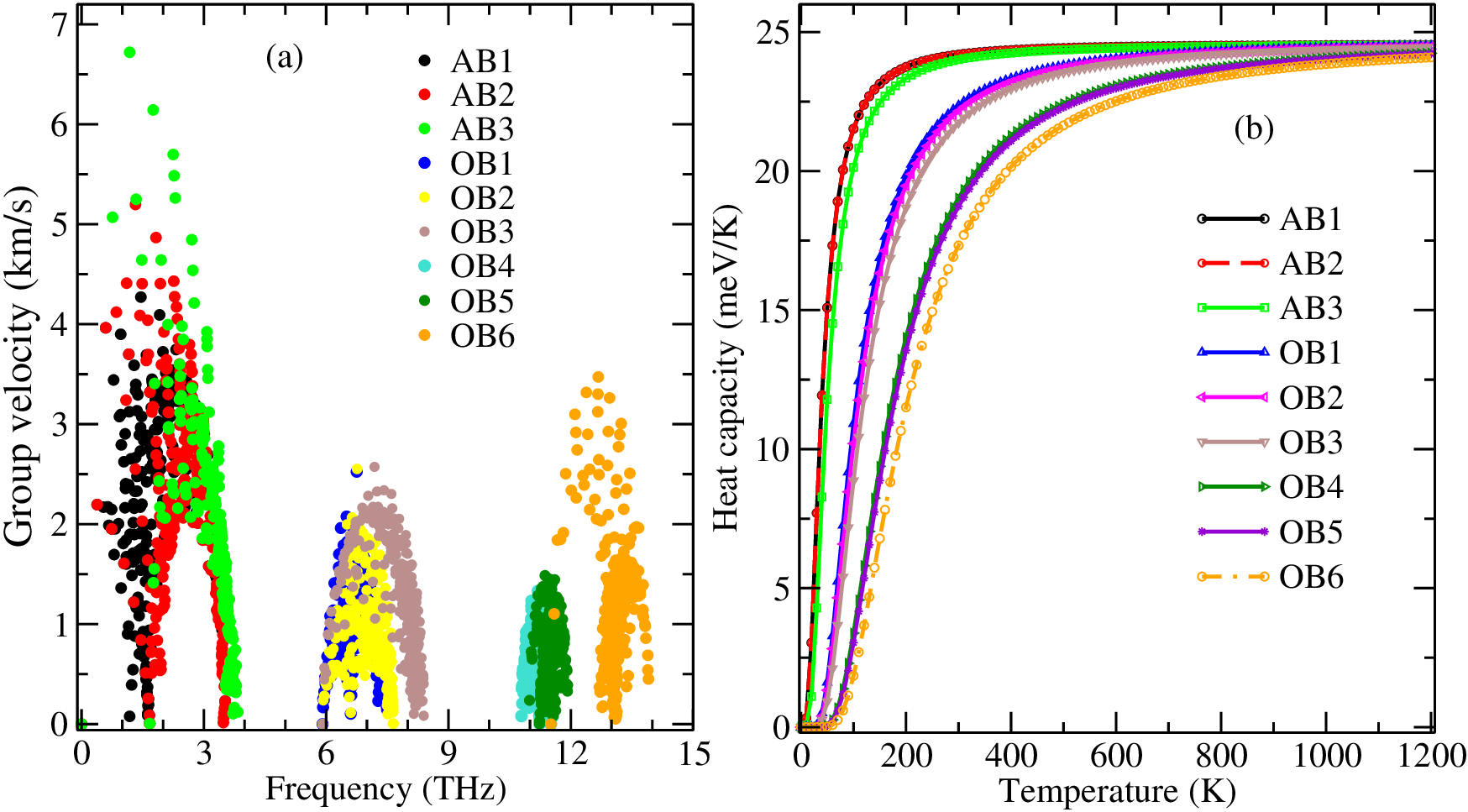} 
\caption{The calculated mode dependent phonon (a) group velocity $\textbf{v}_{\lambda}$ and (b) heat capacity $C_{\lambda}$ for nine phonon branches.} 
\end{figure}

\begin{figure}
\includegraphics[width=7.9cm, height=3.7cm]{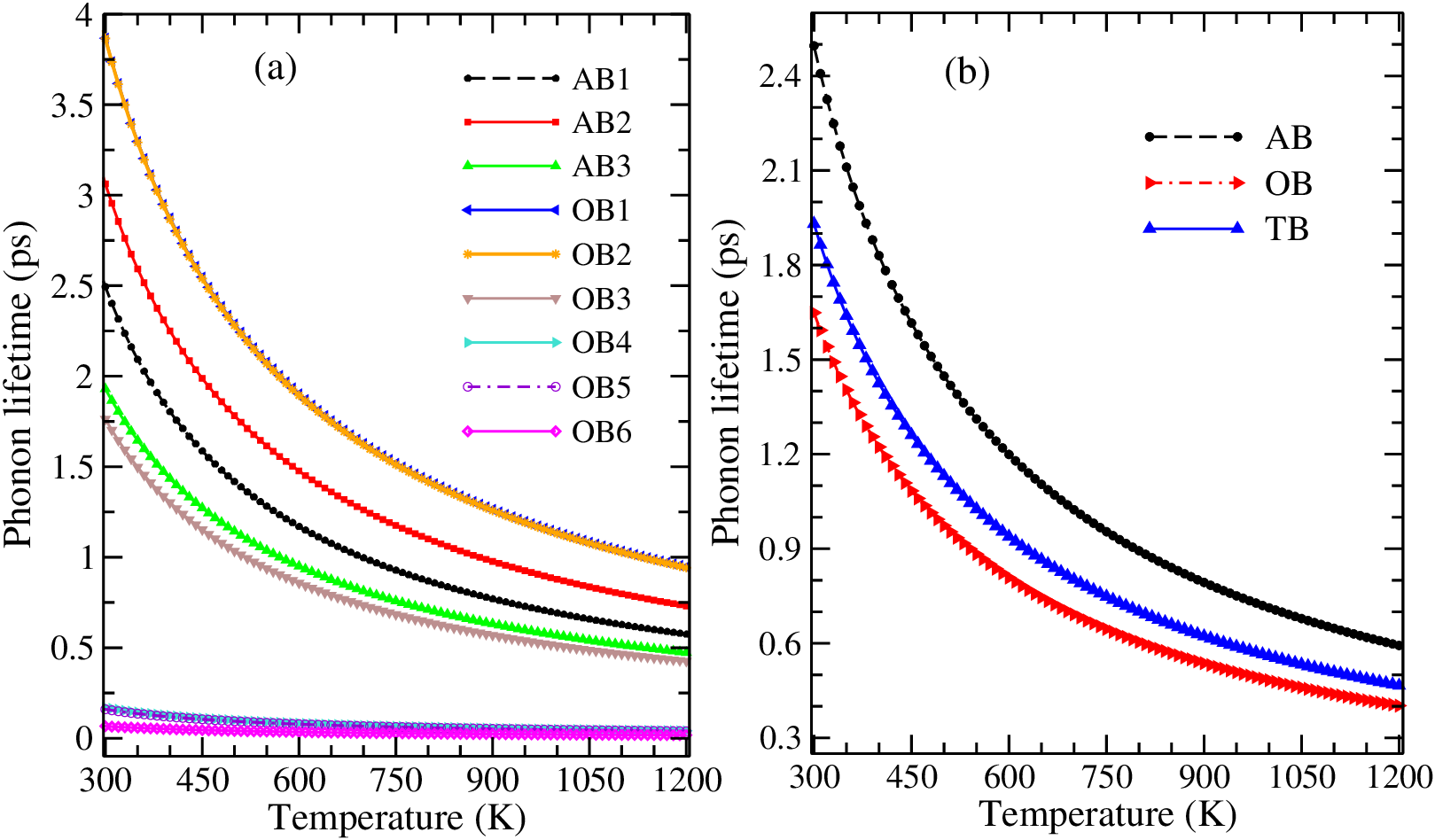} 
\caption{The phonon lifetime $\tau_{\lambda}$ for (a) nine phonon branches (b) acoustic branches (AB), optical branches (OB), and total branches (TB) as a function of temperature.}
\end{figure}

The consideration of temperature-independent $\textbf{v}_{\lambda}$ and nearly non-varying $C_{\lambda}$ behaviour in the temperature range of 300-1200 K motivates us to look closely the temperature variation of $\tau_{\lambda}$ for all phonon branches. Here, from Eq. (4), $\tau_{\lambda}$ due to only PPI is estimated to understand the behaviour of $\kappa_{ph}$ in the 300-1200 K temperature range, which is shown in Fig. 6. In Fig. 6(a), $\tau_{\lambda}$ of all nine branches is obtained by taking the average weight of all q-points in the IBZ. Here, the same method is used for the calculation of $\tau_{\lambda}$ as employed in earlier work by Shastri \textit{et. al} \cite{shastri2021studying}. In the figure, one can clearly notice that OB1 and OB2 have the highest $\tau_{\lambda}$, signifying the lowest phonon-phonon scattering among all the branches. The value of these branches is found to be $\sim$3.9(0.95) ps at 300(1200) K. The last OB9 branch has a shorter $\tau_{\lambda}$ with a value of $\sim$0.07(0.02) ps at 300(1200) K, indicating a higher scattering rate than other phonon branches. All of the branches show decreasing $\tau_{\lambda}$ with increasing temperature, indicating that phonons feel more scattering at high temperatures than phonons at low temperatures. It is also clear from the figure that the decrement rate of $\tau_{\lambda}$ for the last three OB4-OB6 is much slower than for other phonons with increasing temperature that produce low $\kappa_{ph}$. This can be understood through the PDOS plot in Fig. 2, in which the phonon number densities of the last three OB are comparably larger than other branches. The corresponding region contains strong phonon-phonon scattering interactions. The $\tau_{\lambda}$ of TB, AB, and OB is also estimated by averaging the corresponding number of phonon branches, as shown in Fig. 6(b). The $\tau_{\lambda}$ of TB due to PPI is obtained by taking the average of all phonon branches. The figure presents that OB have lower $\tau_{\lambda}$ than AB and the values are $\sim$1.65(2.5) ps and $\sim$0.4(0.6) ps for OB(AB) branches at 300 and 1200 K, respectively. This means that the AB transport more heat energy than OB in the form of $\kappa_{ph}$. The calculated room temperature value of $\tau_{\lambda}$ of TB is $\sim$1.93 ps, while it decreases as temperature increases, with a value of $\sim$0.46 ps at 1200 K. 

\begin{table}
\caption{\small{The variation of $\tau_{\lambda}$ ($\kappa_{ph}$) with a relation of $A_{\tau}$T$^{-x_{\tau}}$ ($A_{\kappa}$T$^{-x_{\kappa}}$) \cite{ashcroft1976} for acoustic branches (AB), optical branches (OB), and total branches (TB).}}
\resizebox{0.43\textwidth}{!}{%
\setlength{\tabcolsep}{1pt}
\begin{tabular}{@{\extracolsep{\fill}}c c c c c c c c c} 
\hline\hline 

Phonon branches &&\multicolumn{1}{c}{$A_{\tau}$(ps$\times$K)}  && \multicolumn{1}{c}{$x_{\tau}$} && \multicolumn{1}{c}{$A_{\kappa}$ (W/m)} && \multicolumn{1}{c}{$x_{\kappa}$}\\ 
     
 \hline
AB1  && 1050  && 1.06   && 786  && 1.05 \\
AB2  && 1113  && 1.03   && 893  && 1.03\\
AB3  && 628   && 1.01   && 463  && 1.02 \\
OB1  && 1263  && 1.02   && 95   && 0.95 \\
OB2  && 1288  && 1.02   && 119  && 0.94 \\
OB3  && 609   && 1.03   && 127  && 0.95\\
OB4  && 59    && 1.02   && 0.81 && 0.85 \\
OB5  && 54    && 1.02   && 1.12 && 0.84 \\
OB6  && 17    && 0.98   && 0.76 && 0.75 \\
\hline
AB  && 924  && 1.04  && 2133 && 1.04 \\
OB  && 548  && 1.02  && 342  && 0.95 \\
TB  && 672  && 1.03  && 2413 && 1.02 \\

\hline\hline 
\end{tabular}}
\end{table}

The total number of phonons in a solid is proportional to temperature T at higher temperatures (T$\gg\Theta_{D}$), which can be understood from Eq. (3). Since a phonon that contributes to thermal conduction is more likely to be scattered with other phonons that are present in crystal, one should expect $\tau_{\lambda}$ to exhibit a decreasing behaviour as temperature rises. At high temperatures, the $C_{\lambda}$ follows the Dulong-Petit law and becomes temperature-independent, which can also be observed in Fig. 5(b). $\textbf{v}_{\lambda}$ is assumed to be a temperature-independent constant in the Debye model, and even in more accurate models, it will not have a significant contribution to $\kappa_{ph}$ \cite{ashcroft1976}. Therefore, in the high-temperature regime, the $\kappa_{ph}$ should decrease as the temperature rises. This temperature-dependence behaviour is confirmed by the experiment, and the rate of decline is $A_{\kappa}$T$^{-x_{\kappa}}$ \cite{ashcroft1976}. The variables $A_{\kappa}$ and $x_{\kappa}$ are temperature-independent and the value of $x_{\kappa}$ lies between 1 and 2 \cite{ashcroft1976}. $A_{\kappa}$ and $x_{\kappa}$ are calculated by fitting the above relation in the respective $\kappa_{ph}$ curve, which is shown in Table 1. The calculated value of $x_{\kappa}$ for TB is $\sim$1.02, which is within the experimental range. From the above discussion, it is important to note that this kind of behaviour is valid for T$\gg\Theta_{D}$. However, in our case, $\Theta_{D}$ is estimated to be $\sim$630 K \cite{solet2022first}, but the calculated $\kappa_{ph}$ follows this trend at temperatures above 300 K. The $x_{\kappa}$ value for each AB is greater than one and less than one for each OB. Also, the $x_{\kappa}$ value for AB(OB) is $\sim$1.04(0.95), indicating that AB controls the behaviour of $x_{\kappa}$ in TB. To deeply understand the $\kappa_{ph}$ behaviour with temperature, the equation $\tau_{\lambda}$ = $A_{\tau}$T$^{-x_{\tau}}$ is also fitted in the respective plots. Here, $A_{\tau}$ and $x_{\tau}$ are also temperature-independent variables. The value of $x_{\tau}$ for TB is $\sim$1.03, while the values for AB and OB are $\sim$1.04 and $\sim$1.02, respectively. Each phonon branch (except OB6) has a $x_{\tau}$ value greater than one, while OB6 has a value of $\sim$0.98. Also, $x_{\tau}$ and $x_{\kappa}$ due to AB1-AB3 are nearly identical, whereas these values differ slightly for OB1-OB6. This means that $x_{\tau}$ values for all AB contribute significantly more to $x_{\kappa}$ than the $x_{\tau}$ values of all OB. The OB2 has highest $A_{\tau}$ of $\sim$1288 ps$\times$K, whereas AB2 has highest $A_{\kappa}$ of $\sim$893 W/m. Finally, the conclusion can be drawn that acoustic modes are the dominant heat carriers in the temperature-dependent $\kappa_{ph}$. 

As a result, AB are dominant carriers in all three previously studied parameters ($\textbf{v}_{\lambda}$, $C_{\lambda}$ and $\tau_{\lambda}$), becoming the larger contributor to the heat transfer process in ScAgC in terms of $\kappa_{ph}$. Among threes, the reduced trend of $\tau_{\lambda}$ with increasing temperature leads directly to the gradual decrement of $\kappa_{ph}$ in Fig. 4 with temperature. One can get a high $ZT$ by reducing the $\kappa_{ph}$ as much as possible. Since $\tau_{\lambda}$ is higher for AB than OB, one can reduce the $\tau_{\lambda}$ of AB by accounting for the extra scattering centres in the form of alloying, nanostructuring, and $etc.$, \cite{snyder2008complex,singh2008nanostructuring} in the energy range of AB. These extra scatterings can provide a low $\kappa_{ph}$ and hence may give rise to a high $ZT$, which is a positive sign for compounds used in TE applications. 

\section{Conclusions}
Here, we have performed the DFT calculations to understand the lattice transport mechanisms of ScAgC combined with the HA and QHA. The phonon band dispersion and phonon DOS are calculated under HA. The obtained positive frequencies (with and without NAC) of dispersion suggest the mechanical stability of ScAgC in the FCC structure. The value of $\alpha(T)$ is found to be $\sim$4$\times$$10^{-6}$ K$^{-1}$ and $\sim$4.6$\times$$10^{-6}$ K$^{-1}$ at 300 K and 1200 K, respectively. Similarly, first-principles based anharmonic phonon calculations have been used to analyse the $\kappa_{ph}$ and the value is obtained as $\sim$7.4(1.8) Wm$^{-1}$K$^{-1}$ at 300(1200) K. The value of total $\tau_{\lambda}$ at 300 K is calculated to be $\sim$1.93 ps, whereas it is observed as $\sim$0.46 ps at 1200 K. The highest calculated $\textbf{v}_{\lambda}$ for AB (OB) is $\sim$6.7 (3.5) km/s. At 200 K, the AB (OB) have the highest (lowest) $C_{\lambda}$ with a value of $\sim$23.5 (11.5) meV/K. By fitting the equation of $A_{\kappa}$T$^{-x_{\kappa}}$($A_{\tau}$T$^{-x_{\tau}}$) in $\kappa_{ph}$($\tau_{\lambda}$) curve, the temperature-dependent behaviour is also understood. The $x_{\kappa}$ value for AB(OB) is predicted to be $\sim$1.04(0.95), while it is $\sim$1.02 for TB. Similarly, the obtained value of $x_{\tau}$ is $\sim$1.04(AB), $\sim$1.02(OB), and $\sim$1.03(TB). Our study can be helpful in order to use ScAgC for renewable energy sources such as TE and PV applications. \\

\textbf{Author contribution statement} \\

This problem is solely formulated by S.K.P. Under his guidance, all the numerical calculations, careful study of data and preparing the manuscript are done by V.K.S. After the discussion between both of V.K.S. and S.K.P., the results and the comments on manuscript at each stages of the revision are prepared. \\

\textbf{Data availability statement} \\

The computational data of this manuscript will be made available on reasonable request.


\bibliography{ref}
\bibliographystyle{apsrev4-2}

\end{document}